\input harvmac
\noblackbox

\input epsf

\newcount\figno
\figno=0
\def\fig#1#2#3{
\par\begingroup\parindent=0pt\leftskip=1cm\rightskip=1cm\parindent=0pt
\baselineskip=11pt
\global\advance\figno by 1
\midinsert
\epsfxsize=#3
\centerline{\epsfbox{#2}}
\vskip 12pt
{\bf Fig.\ \the\figno: } #1\par
\endinsert\endgroup\par
}
\def\figlabel#1{\xdef#1{\the\figno}}
\def\encadremath#1{\vbox{\hrule\hbox{\vrule\kern8pt\vbox{\kern8pt
\hbox{$\displaystyle #1$}\kern8pt}
\kern8pt\vrule}\hrule}}

\def\apm{{\alpha^{\prime}}}


\def\ep{{\epsilon}}
\def\b{{\beta}}
\def\p{\partial}

\def\eqn#1#2{\xdef #1{(\secsym\the\meqno)}\writedef{#1\leftbracket#1}%
\global\advance\meqno by1$$#2\eqno#1\eqlabeL#1$$}

\lref\BrusteinPY{
R.~Brustein and K.~Roland,
``Space-time versus world sheet renormalization group equation in string theory,''
Nucl.\ Phys.\ B {\bf 372}, 201 (1992).
}

\lref\ashokehet{
A.~Sen,
``Equations Of Motion For The Heterotic String Theory From The Conformal Invariance Of The Sigma Model,''
Phys.\ Rev.\ Lett.\  {\bf 55}, 1846 (1985).
}

\lref\LovelaceKR{
C.~Lovelace,
``Stability of string vacua. 1. A new picture of the renormalization group,''
Nucl.\ Phys.\ B {\bf 273}, 413 (1986).
}

\lref\LovelaceYV{
C.~Lovelace,
``Strings in curved space,''
Phys.\ Lett.\ B {\bf 135}, 75 (1984).
}

\lref\BanksQS{
T.~Banks and E.~J.~Martinec,
``The renormalization group and string field theory,''
Nucl.\ Phys.\ B {\bf 294}, 733 (1987).
}

\lref\DowkerGB{ F.~Dowker, J.~P.~Gauntlett, G.~W.~Gibbons and
G.~T.~Horowitz, ``The decay of magnetic fields in Kaluza-Klein
theory,'' Phys.\ Rev.\ D {\bf 52}, 6929 (1995)
[arXiv:hep-th/9507143].
}

\lref\DowkerSG{ F.~Dowker, J.~P.~Gauntlett, G.~W.~Gibbons and
G.~T.~Horowitz, ``Nucleation of $P$-Branes and fundamental
Strings,'' Phys.\ Rev.\ D {\bf 53}, 7115 (1996)
[arXiv:hep-th/9512154].
}

\lref\CallanIA{ C.~G.~Callan, E.~J.~Martinec, M.~J.~Perry and
D.~Friedan, ``Strings in background fields,'' Nucl.\ Phys.\ B 
{\bf 262}, 593 (1985).
}

\lref\FradkinPQ{ E.~S.~Fradkin and A.~A.~Tseytlin, ``Effective
field theory from quantized strings,'' Phys.\ Lett.\ B {\bf 158},
316 (1985).
}

\lref\ZamolodchikovGT{ A.~B.~Zamolodchikov, ``'Irreversibility' of
the flux of the renormalization group in a 2-D field theory,''
JETP Lett.\  {\bf 43}, 730 (1986) [Pisma Zh.\ Eksp.\ Teor.\ Fiz.\
{\bf 43}, 565 (1986)].
}

\lref\AffleckTK{ I.~Affleck and A.~W.~Ludwig, ``Universal
noninteger 'ground state degeneracy' in critical quantum
systems,'' Phys.\ Rev.\ Lett.\  {\bf 67}, 161 (1991).
}

\lref\HarveyNA{ J.~A.~Harvey, D.~Kutasov and E.~J.~Martinec, ``On
the relevance of tachyons,'' arXiv:hep-th/0003101.
}

\lref\HarveyGQ{ J.~A.~Harvey, S.~Kachru, G.~W.~Moore and
E.~Silverstein, ``Tension is dimension,'' JHEP {\bf 0003}, 001
(2000) [arXiv:hep-th/9909072].
}

\lref\SenSM{ A.~Sen, ``Tachyon condensation on the brane antibrane
system,'' JHEP {\bf 9808}, 012 (1998) [arXiv:hep-th/9805170].
}

\lref\TseytlinWW{ A.~A.~Tseytlin, ``Renormalization of M\"obius
infinities and partition function representation for string theory
effective action,'' Phys.\ Lett.\ B {\bf 202}, 81 (1988).
}

\lref\TseytlinTV{ A.~A.~Tseytlin, ``M\"obius infinity subtraction
and effective action in sigma model approach to closed string
theory,'' Phys.\ Lett.\ B {\bf 208}, 221 (1988).
}

\lref\DabholkarWN{ A.~Dabholkar and C.~Vafa, ``tt* geometry and
closed string tachyon potential,'' JHEP {\bf 0202}, 008 (2002)
[arXiv:hep-th/0111155].
}

\lref\VafaRA{C.~Vafa,
``Mirror symmetry and closed string tachyon condensation,''
arXiv:hep-th/0111051.
}

\lref\bondir{H.~Bondi, M.~G.~J.~Van der Burg and A.~W.~K.~Metzner,
``Gravitational waves in general relativity VII. Waves from 
axi-symmetric isolated systems'', Proc.\ Roy.\ Soc.\ London,
Ser.\ {\bf  A 269}, 21 (1962). }

\lref\sachs{R.~K.~Sachs,``Gravitational waves in general
relativity VIII. Waves in asymptotically flat space-time'', 
Proc.\ Roy.\ Soc.\ London, Ser.\ {\bf A 270}, 103 (1962).}

\lref\adm{R.~Arnowitt, S.~Deser, and C.~W.~Misner, ``The
dynamics of general relativity'', in Witten, Louis, ed.,
Gravitation: An Introduction to Current Research, 227-265, 
(Wiley, New York, 1962). }

\lref\polchinski{J.~Polchinski, ``String Theory, Volume II'',
Cambridge University Press.}

\lref\wald{R.~Wald, ``General Relativity'', University of 
Chicago Press}

\lref\tseytlin{
A.~A.~Tseytlin,
``Conditions of Weyl invariance of two-dimensional sigma model 
from equations of stationarity of 'central charge' action,''
Phys.\ Lett.\ B {\bf 194}, 63 (1987).
}

\lref\chicago{
J.~A.~Harvey, D.~Kutasov, E.~J.~Martinec and G.~Moore,
``Localized tachyons and RG flows,'' arXiv:hep-th/0111154.
}

\lref\hh{
S.~W.~Hawking and G.~T.~Horowitz,
``The gravitational Hamiltonian, action, entropy and surface terms,''
Class.\ Quant.\ Grav.\  {\bf 13}, 1487 (1996)
[arXiv:gr-qc/9501014].
}

\lref\CecottiTH{
S.~Cecotti and C.~Vafa,
``Massive orbifolds,''
Mod.\ Phys.\ Lett.\ A {\bf 7}, 1715 (1992)
[arXiv:hep-th/9203066].
}

\lref\sw{N.~Seiberg and E.~Witten, ``String Theory and 
Noncommutative Geometry'', hep-th/9912072.} 

\lref\berg{E.~Bergshoeff,
D.~S.~Berman, J.~P.~van der Schaar and P.~Sundell, ``A
noncommutative M-theory five-brane,'' hep-th/0005026.}

\lref\david{
J.~R.~David,
``Unstable magnetic fluxes in heterotic string theory,''
arXiv:hep-th/0208011.
}

\lref\dhgm{
J.~R.~David, M.~Gutperle, M.~Headrick and S.~Minwalla,
``Closed string tachyon condensation on twisted circles,''
JHEP {\bf 0202}, 041 (2002)
[arXiv:hep-th/0111212].
}

\lref\dv{
A.~Dabholkar and C.~Vafa,
``tt* geometry and closed string tachyon potential,''
JHEP {\bf 0202}, 008 (2002)
[arXiv:hep-th/0111155].
}

\lref\aps{
A.~Adams, J.~Polchinski and E.~Silverstein,
``Don't panic! Closed string tachyons in ALE space-times,''
JHEP {\bf 0110}, 029 (2001)
[arXiv:hep-th/0108075].
}

\lref\kraus{
P.~Kraus, A.~Ryzhov and M.~Shigemori,
``Strings in noncompact spacetimes: Boundary terms and conserved  charges,''
arXiv:hep-th/0206080.
}

\lref\CurciHI{
G.~Curci and G.~Paffuti,
``Consistency Between The String Background Field Equation Of Motion And The Vanishing Of The Conformal Anomaly,''
Nucl.\ Phys.\ B {\bf 286}, 399 (1987).
}

\lref\TseytlinTT{
A.~A.~Tseytlin,
``Conformal Anomaly In Two-Dimensional Sigma Model On Curved Background 
And Strings,''
Phys.\ Lett.\  {\bf 178B}, 34 (1986).
}

\lref\tseytlinreview{
A.~A.~Tseytlin,
``On The Renormalization Group Approach To String Equations Of Motion,''
Int.\ J.\ Mod.\ Phys.\ A {\bf 4}, 4249 (1989).}

\lref\CallanJB{
C.~G.~Callan, I.~R.~Klebanov and M.~J.~Perry,
``String Theory Effective Actions,''
Nucl.\ Phys.\ B {\bf 278}, 78 (1986).
}

\lref\ShoreHK{
G.~M.~Shore,
``A Local Renormalization Group Equation, Diffeomorphisms, And Conformal Invariance In Sigma Models,''
Nucl.\ Phys.\ B {\bf 286}, 349 (1987).
}

\lref\DasDA{
S.~R.~Das, A.~Dhar and S.~R.~Wadia,
``Critical Behavior In Two-Dimensional Quantum Gravity And Equations Of Motion Of The String,''
Mod.\ Phys.\ Lett.\ A {\bf 5}, 799 (1990).
}

\lref\aflu{I.~Affleck and A.~W.~Ludwig, ``Exact conformal field theory
results on the multichannel Kondo effect: Single-fermion
Green's function, self-energy, and resistivity,
Phys.\ Rev.\  {\bf B48}, 7297 (19yy).
}

\lref\pol{
J.~Polchinski,
``Scale And Conformal Invariance In Quantum Field Theory,''
Nucl.\ Phys.\ B {\bf 303}, 226 (1988).
}

\lref\cp{
G.~Curci and G.~Paffuti,
``Infrared Problems In Two-Dimensional Generalized Sigma Models,''
Nucl.\ Phys.\ B {\bf 312}, 227 (1989).
}

\Title {\vbox{ \baselineskip12pt
\hbox{hep-th/0211063}\hbox{HUTP-02/A058}\hbox{SU-ITP-02/41} }}
{\vbox{ \centerline{Spacetime Energy Decreases under} \bigskip
\centerline{World-sheet RG Flow}\centerline{} }}

\centerline{ Michael ${\rm Gutperle,}^a$\foot{On leave of absence
from Department of Physics and Astronomy, UCLA, Los Angeles CA.}
Matthew ${\rm Headrick,}^b$ Shiraz ${\rm Minwalla,}^b$ and
Volker ${\rm Schomerus}^c$}

\bigskip
\centerline{$^a$ Department of Physics, Stanford University,
Stanford CA 94305, USA}
\smallskip

\centerline{$^b$ Jefferson Physical Laboratory, Harvard
University, Cambridge MA 02138, USA}
\smallskip

\centerline{$^c$ Service de Physique Th\'eorique, CEA - Saclay, 
F - 91191 Gif-sur-Yvette Cedex, France}
\smallskip

\vskip .3in \centerline{\bf Abstract} {We study renormalization
group flows in unitary two dimensional sigma models with
asymptotically flat target spaces. Applying an infrared cutoff to
the target space, we use the Zamolodchikov $c$-theorem to
demonstrate that the target space ADM energy of the UV fixed point
is greater than that of the IR fixed point: spacetime energy
decreases under world-sheet RG flow. This result mirrors the well
understood decrease of spacetime Bondi energy in the time
evolution process of tachyon condensation.}
\smallskip
\Date{}
\listtoc
\writetoc

\newsec{Introduction}

Perturbative string theory establishes several intriguing
connections between world-sheet and spacetime dynamics. As is 
well known, consistency of classical string propagation requires
the world-sheet sigma model to be conformal, which in turn imposes
$\apm$ corrected Einstein equations on spacetime fields 
\refs{\LovelaceYV,
 \ashokehet, \CallanIA, \FradkinPQ,\LovelaceKR,\CurciHI, 
\TseytlinTT, \CallanJB, 
\ShoreHK}. The spectrum of 
fluctuations about solutions to the spacetime equations is 
in one-to-one correspondence with marginal operators of the 
corresponding world-sheet CFT, and the scattering of these 
fluctuation modes is computed by world-sheet correlators of 
the corresponding marginal operators. In this paper we will be 
concerned with yet another aspect of perturbative string
dynamics, namely the connection between time evolution 
of the background and world-sheet renormalization group (RG) 
flows. The idea that these processes are related has been around for a 
long time, and has occasionally been employed in practical computations, 
most notably in the recent studies of localized closed string tachyon 
condensation \refs{\aps,\VafaRA,\dv,\chicago,\dhgm,\david} that have motivated
the present work. Nonetheless the existing evidence for a definite 
relationship between these two processes  appears rather scarce. 
Our aim in this paper is to present a non-trivial 
consistency check of the conjectured relationship between time evolution
and world-sheet RG flow. 

This conjectured relationship springs from the following observation:
Consider a type II string background of the form CFT$_1+\rm{CFT}_2$, 
where CFT$_2$ is any unitary conformal field theory with ${\hat c}=d$, 
and CFT$_1$ is the free sigma model on $R^{9-d,1}$ with fields 
$X^a$ ($a=0,\ldots,9-d$). It is not difficult to see that the  
operator spectrum of CFT$_2$ determines the stability of this 
background under time evolution. In fact, a conformal operator 
$O$ of dimension $(\delta, \delta)$ in CFT$_2$ may be combined 
with a momentum factor from CFT$_1$ to yield a marginal operator 
in the full CFT: ${\hat O}=e^{iP\cdot X}O$ where $\apm P^2/2+
\delta=1$. Thus the world-sheet operator ${\hat O}$ corresponds 
to a spacetime fluctuation with squared mass $M^2=2(\delta-1)
/\apm$ and so represents a tachyon or instability when $O$ is 
a relevant operator. We therefore conclude that the existence of 
relevant operators in the spectrum of CFT$_2$ implies 
an instability of the corresponding string background.

It is tempting to go beyond such an `infinitesimal'
statement and to conjecture a relation between the full dynamical 
evolution in string theory and renormalization group flows 
on the world-sheet. Clearly, the two sides have many features 
in common.
A world-sheet RG flow away from an unstable string background
ends at an infrared conformal field theory that may generically 
be expected to be stable. The IR fixed point can be a theory with
 a mass gap, corresponding to a flow toward a noncritical string theory and
 will not be considered in this note. Similarly, after all the dust has 
settled, the dynamical process of tachyon condensation is 
generically expected to decay into a stable solution of string 
theory. It has often been conjectured that the endpoints of 
these two processes are the same (see e.g.\ 
\refs{\BanksQS, \BrusteinPY, \DasDA}).
In our attempt to provide evidence for such a conjecture, 
we shall mainly focus on a special class of string backgrounds
for which CFT$_2$ is a sigma model onto an {\it asymptotically 
flat} $d$ dimensional space. We also restrict attention to 
condensation processes throughout which spacetime remains
asymptotically flat. This requires, in particular, that the
relevant operators of CFT$_2$ are {\it localized} in target 
space. These assumptions will be relaxed slightly toward the 
end of the paper. 

Intuitively, the decay of an unstable string background is a
directional process: condensation processes are driven by the 
desire of a system to minimize its energy. Within the setup 
we specified in the previous paragraph, it is possible to make 
this intuition precise
in terms of the so-called Bondi energy of asymptotically flat spacetimes,
which can be considered as the difference 
between the total energy and the energy that has escaped 
away as radiation (see section 2.1 for a more precise formulation).
Under appropriate assumptions it can 
be shown that the Bondi energy
\refs{\bondir,\sachs} of a gravitational background decreases 
in time. In particular, if a gravitational background 
decays from a initial configuration toward a static solution, 
then the energy of the latter is smaller than the energy 
of the unstable solution in the far past. These mathematical 
results on Bondi energy feed the expectation that condensation 
processes in our string backgrounds proceed so as to lower some 
`stringy energy.' Even though we do not know the precise 
definition of the latter, we expect it to agree with the 
relativists' notion of energy whenever the gravity 
approximation is valid. 

This brings us to a description of our main results. 
Based on the arguments we have sketched above we shall take 
the directionality of stringy time evolution for granted, and 
show that the same behavior is found for the corresponding RG 
flows. More precisely, we will define some quantity on the 
world-sheet which decreases along the flows and relate it to 
gravitational energies.  

At first sight, this seems to be a rather simple task, since 
Zamolodchikov's famous $c$-theorem \ZamolodchikovGT\ asserts
that the  $C$ function, an off-shell generalization of the 
central charge, is a (not necessarily strictly) decreasing 
function of RG scale. But for our non-compact backgrounds, 
the $C$ function is completely determined by the asymptotics 
of target space and so remains constant along the flow. 
Nonetheless, by applying an infrared cutoff (compactification)
to the target space, we will be able to employ Zamolodchikov's 
$c$-theorem and establish a directionality for RG flows in 
CFT$_2$. The cutoff in target space will allow us to remove 
the leading universal contribution from the $C$-function and 
to prepare a functional $S$ that evolves from a higher value 
in the ultraviolet (UV) to a lower value at the infrared (IR)
 fixed point. We shall 
then show that the change in $S$ in RG flow from the UV to the 
IR is precisely the difference between the ADM energies of the 
UV and IR target spaces. Thus, whenever quantities in
the UV and IR limits can be well approximated by gravity computations, 
our results demonstrate that the ADM energy for the IR fixed 
point of our RG flows is lower than the ADM energy of the 
unstable UV fixed point. Using the known relation between 
ADM and Bondi energy (see also section 2.1 below), we therefore 
obtain a perfect agreement with the analogous result from the 
target space analysis that we have sketched above. Toward the 
end of this note we also present a generalization of our 
analysis to flows on non-compact sigma models with arbitrary but 
fixed asymptotics, and to flows on asymptotically conical 
two-dimensional sigma models that interpolate between spacetimes 
with different deficit angles.

We emphasize that CFT$_1$, which contains the time direction, is a mere
spectator throughout the RG flow of CFT$_2$, whose unitarity justifies the
use of the $c$-theorem. This may raise the following question in the mind
of the reader: since the energy of a gravitational solution is roughly
measured by the falloff near infinity of the time-time component of the
metric, isn't the energy identically zero for all solutions we consider?
The answer is that the physically relevant energy is determined by the
Einstein frame metric,\foot{More precisely, due to the translational
invariance in the $9-d$ spatial directions of CFT$_1$, it is the $d+1$
dimensional ADM energy that is properly defined (conceptually it is perhaps
best to regard the other directions as compactified). Hence it is the $d+1$
dimensional Einstein frame metric we will employ.} not the string
frame metric; thanks to the evolution 
of the dilaton, the time-time component of the Einstein frame metric will
generically undergo quite non-trivial evolution under RG flow (although
solutions such as non-dilatonic black holes, for example, would be excluded
from our analysis). In the body of this paper, we will distinguish between
the two metrics by denoting the string frame metric 
$G_{\mu\nu}$ and the Einstein frame one $g_{\mu\nu}$.

\newsec{Essential Background}

The purpose of this section is to collect a number of well 
known results that provide much of the technical background 
for our analysis. In subsection 2.1 we recall the notion of 
Bondi energy in General Relativity, and we describe its monotonic 
decrease in light-cone time as well as its relation to the ADM 
energy. Subsection 2.2 contains a short review of the $c$-theorem. 
In subsection 2.3 we list some results from Tseytlin's computation 
\tseytlin\ of the $C$ function in $\apm$ perturbation theory.

\subsec{Energy in asymptotically flat spacetimes}

Relativists have defined at least two useful notions of
energy in asymptotically flat spacetimes: ADM energy \adm\ 
and Bondi energy \refs{\bondir,\sachs}. These energies are both completely
determined by the behavior of the spacetime metric `at infinity';
they differ in {\it where} at infinity they are measured.

The ADM energy of a $d+1$ dimensional asymptotically flat
spacetime is determined by the behavior of its metric on 
an infinite $d-1$ dimensional sphere $i^0$ that constitutes 
the boundary of space at any constant time. Explicitly,%
\foot{In our conventions the Einstein action is $S={1\over
2\kappa^2}\int \sqrt{-g} R$. }
\eqn\energy{
E_{\rm{ADM}} =
{1\over 2\kappa^2 } 
\int_{i^0} dS^j \left(\p_i h_{ij}- \p_j h_{ii} \right)
,}
where $h_{ij}=g_{ij}-\delta_{ij}$ and the 
orientation of $i^0$ is chosen such that the normal vector 
points to the outside ($i,j$ are spatial indices).
It is not difficult to see that ADM energy is conserved under 
the gravitational time evolution.

Bondi energy, on the other hand, is determined by the behavior 
of the spacetime metric on a $d-1$ dimensional sphere of radius 
$r$ and at time $t$ (centered about the origin of space), where 
$r \to \infty$, $t \to \infty$  with $t-r= \lambda$ fixed. In 
other words, Bondi energy is evaluated at $\CI^+$ on a Penrose 
diagram. The light-cone time $\lambda$ labels where on $\CI^+$ 
this energy is measured (see fig.\ 1). An explicit formula for 
Bondi energy as an integral of a function of the metric over 
a particular sphere $S^{d-1}$ on $\CI^+$ may be found, for 
instance, in \wald, but will not be needed thoughout here. 
\fig{ADM energy, the energy
contained in the largest spatial slice available in an
asymptotically flat space (see the first Penrose diagram above) is
measured by the fall-off of the metric at $i^0$. Bondi energy
represents the energy contained in smaller spatial slices (see the
second Penrose diagram above), measured by the fall-off of the
metric at $\CI^+$. Bondi energy is a decreasing function of
$\lambda$.} {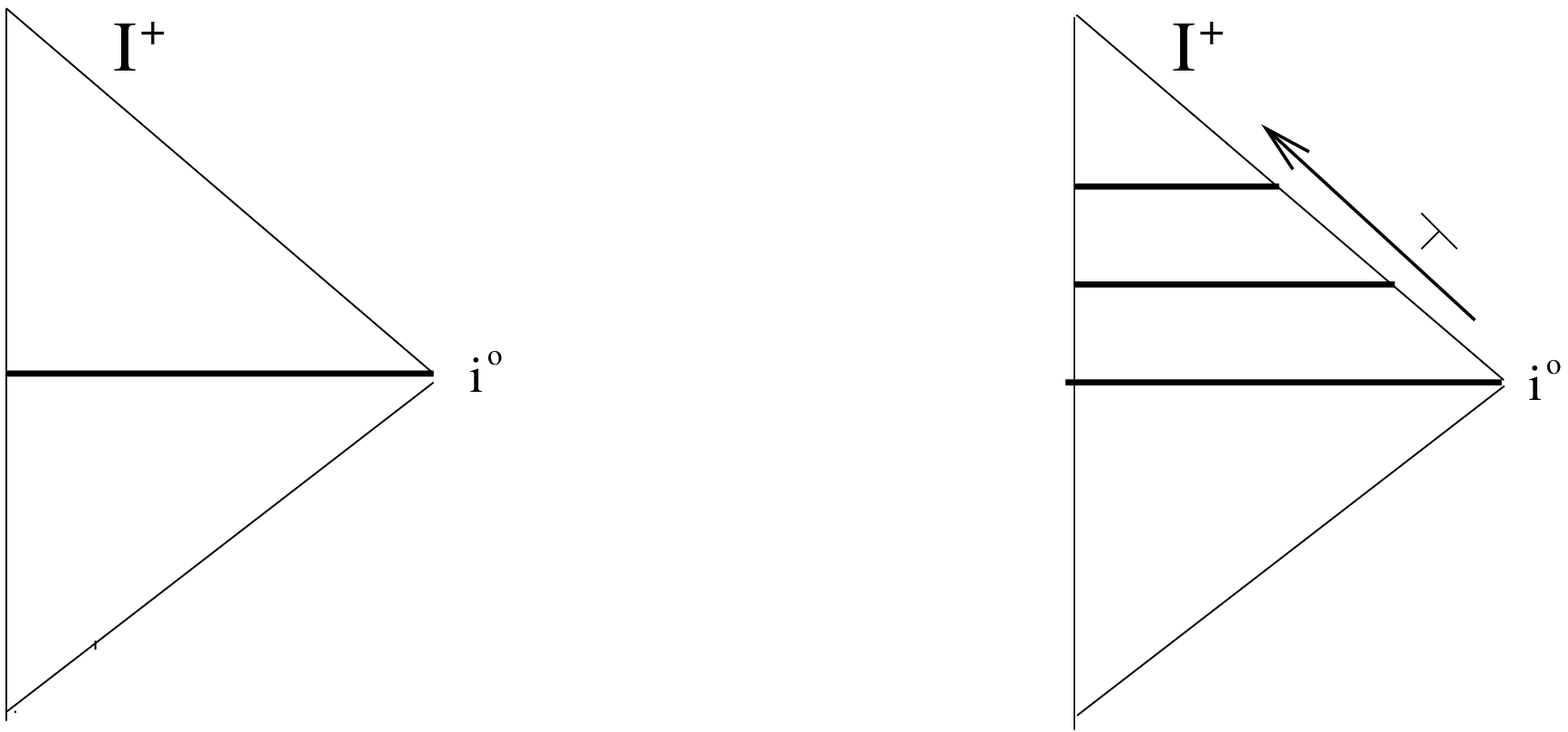}{3.5truein}

Note that Bondi energy is not conserved as a function of $\lambda$; this
and other interesting features can be illustrated 
by the following thought experiment. Consider a universe that 
is empty except for a small bomb of mass $M$ located at the 
origin of space. Let this bomb sit dormant until it explodes at 
$t=0$. In this moment, all its energy is converted into radiation 
that propagates radially away to infinity. Since ADM energy is 
measured on infinite spheres at constant time which are never 
reached by the outgoing radiation, ADM energy of this spacetime 
is $M$, identical to that of a second universe in which the bomb 
never explodes. On the other hand, Bondi energy is determined on a family
of light-cones whose apex 
is located in the origin of space. For $\lambda < 0$, the associated
light-cone lies in a region of spacetime that never receives the news 
about the explosion. Hence, the Bondi energy at these values of 
$\lambda$ is simply $M$. When $\lambda > 0$, on the other hand, 
the Bondi energy vanishes since it is measured on light-cones which
lie entirely within flat space. Thus the Bondi energy of our 
spacetime is a monotonically decreasing function of $\lambda$
which interpolates between the ADM energies of the `initial' 
unstable solution (the bomb) and of the `final' static 
solution (flat space).

The thought experiment of the previous paragraph illustrates several
general properties of Bondi energy, established by relativists in the 
late 1970s (see the end of chapter 11 in \wald\ for a review and
references). At $\lambda = - \infty$, the Bondi energy of a spacetime 
is equal to its ADM energy. In general, Bondi 
energy can also be shown to decrease in $\lambda$ and the difference 
between the Bondi energy at two different light-cone times $\lambda_1$ 
and $\lambda_2$ is equal to the energy lost due to radiation in the 
band $[\lambda_1, \lambda_2] \subset\CI^+$. Finally, if a given 
spacetime agrees with an auxiliary static spacetime in the region 
between two light-cone times $\lambda_1$ and $\lambda_2$, then their 
Bondi energy coincides on the interval $[\lambda_1,\lambda_2]$. 
Obviously, the Bondi energy of any static spacetime is constant 
and equal to its ADM energy. 

As we claimed in the introduction, these general results on Bondi 
energy and its relation to ADM energy bridge between our findings
on RG flows and the general expectations for time evolution in 
string theory. Below we will study RG flows from UV to IR fixed points on 
the world-sheet of the string; we will demonstrate 
that whenever quantities in the UV and IR limits can be well approximated 
by gravity computations, ADM energy of the UV fixed point is larger than 
the ADM energy of the IR fixed point. On the other hand, if we take the 
background fields at the UV fixed point in a region of light-cone time 
$\lambda < \lambda_1$, perturb them slightly at $\lambda \sim \lambda_1$ 
and let the perturbed system evolve in time using the gravity equations of 
motion, the system is expected to settle down into a stable static 
solution plus radiation at some later light-cone time $\lambda_2 > 
\lambda_1$. The Bondi energy of the resulting spacetime is 
$E^{(1)}_{\rm ADM} = E^{\rm UV}_{\rm ADM}$ for  $\lambda < \lambda_1$ 
and it is equal to $E^{(2)}_{\rm ADM}= E^{\rm IR}_{\rm ADM}$ at 
$\lambda > \lambda_2$. Since Bondi energy is known to decrease 
we conclude  $E^{(1)}_{\rm ADM} \geq E^{(2)}_{\rm ADM}$, just 
as for our RG flows on the world-sheet.

\subsec{The Zamolodchikov $c$-theorem}

The purpose of this subsection is to review the Zamolodchikov 
$c$-theorem. Our discussion follows the presentation and 
notation of \polchinski.

In a rotationally invariant Euclidean two-dimensional quantum field
theory, two-point functions of stress energy tensors are constrained to
take the following form
\eqn\fotta{\eqalign{\langle T_{zz}(z)T_{zz}(0) \rangle&={F(r) \over
z^4}, \cr
\langle T_{z{\bar z}}(z)T_{zz}(0) \rangle&={G(r) \over 4 z^3 {\bar
z}}, \cr
\langle T_{z{\bar z}}(z)T_{z\bar{z}}(0) \rangle&=
{H(r) \over 16 z^2 {\bar z}^2}, \cr}}
where $r=\sqrt{z \bar{z}}$.
The numerical factors on the right hand side have been chosen for later
convenience. Ward identities associated with conservation of the stress 
tensor yield
\eqn\totpso{\eqalign{
\bar{\p}\langle T_{zz}(z)T_{zz}(0) \rangle + {\p}\langle
T_{z\bar{z}}(z)T_{zz}(0) \rangle &= 0, \cr
\bar{\p}\langle T_{zz}(z)T_{z{\bar z}}(0) \rangle +
{\p}\langle T_{z\bar{z}}(z)T_{z \bar{z}}(0) \rangle &= 0.
}}
Combining \totpso\ and \fotta\ we find
\eqn\fe{\eqalign{& 4 {\dot F}+{\dot G}-3 {G}=0, \cr
&4 {\dot G}-4 G+{\dot H}-2 H=0.}}
where ${\dot a}= r (da/dr)$. Defining
\eqn\zcf{
C(r) = 2F(r)-G(r)-{3H(r) \over 8},
}
we have
\eqn\zct{
\dot{C}=-{3 \over 4}H.
}
At a conformal point $G$ and $H$ vanish (since $T_{z {\bar z}}$
vanishes in any conformal field theory) while $2F$ reduces to 
$c$, the constant central charge. Consequently $C(r) =c$ in a 
conformal field theory. Away from conformal points, $C(r)$ is 
generically not constant. In fact, the positivity of $H(z)$ in 
a unitary field theory implies that $C(r)$ is a monotonically 
non-increasing function of radius. Equivalently, by the 
Callan-Symanzik equation, $C(r)$ is a pointwise non-increasing 
function along renormalization group flows \foot{This seemingly innocuous
claim requires several qualifications. As was emphasized in \pol\ (see 
also \tseytlinreview), the action for a two dimensional field theory
on a flat world sheet does not uniquely specify the stress tensor for the
theory; the ambiguity in the stress tensor 
corresponds to the freedom in the choice of the dilaton. $C(r)$
decreases pointwise under RG flows only if the ambiguity in the definition 
of the stress tensor is fixed, scale by scale, to insure that 
$T_{\mu\nu}$ obeys 
the naive Callan-Symanzik equation. Such a redefinition of 
the stress tensor is always possible (see \pol); in fact the redefined
stress tensor, as a function of scale may be determined by the requirement that
the dilaton evolve according to its RG equations. This will be the 
prescription employed in the next section for the computation of the $C$
function in perturbation theory. The application of the Callan Symanzik 
equation to $C(r)$ is also complicated by the usual infrared
divergences associated with massless two dimensional scalars \cp\ 
on attempting to expand them about a point in their moduli
space. Presumably these divergences may be dealt with in the usual
fashion (shifting to the correct vacuum, a wave function on moduli
space) and do not present a problem in principle. We would like to thank 
H. Osborn, A. Strominger, A. Tseytlin and especially J. Polchinski for very
illuminating discussions on this topic.}. As we have already 
seen it also reduces to the constant central charge at conformal 
points. These statements are the content of the Zamolodchikov 
$c$-theorem.

To illustrate some special issues in sigma-models with non-compact 
target space it is useful to recast equation \zct\ by inserting the
well known equation for the trace anomaly in a quantum field theory, 
\eqn\sebf{
T_{z \bar{z}} = -\pi\b^a O_a.
} 
Here the $O_a$ form a basis of relevant and marginal operators. We 
find that the $C$ function 
\eqn\zctm{\eqalign{
{\dot C}&=
{-12 \pi^2 z^2 {\bar z}^2} \b^a\b^b \langle O_a(z) O_b(0) \rangle \cr
&= {-12\pi^2} \b^a \b^b{\cal G}_{ab}(z)
}}
can be expressed through the Zamolodchikov metric,  
\eqn\zammet{
{\cal G}_{ab}(z)=z^2{\bar z}^2\langle O_a(z) O_b(0) 
\rangle.
}
and hence through the matrix of two point functions. The latter 
is schematically given by
\eqn\matrixrel{
\langle O_a(0) O_b(z) \rangle =
{\int \CD X^i e^{-S[X]} O_a(X(0))O_b(X(z) ) \over \int \CD X^i 
e^{-S[X]}}.
}
As observed in \refs{\chicago,\CecottiTH}, 
the denominator of the matrix element 
\matrixrel\ is proportional to the volume of the target space 
(from the integral over the zero mode) and hence it is infinite 
for our non-compact 
backgrounds. We will cut off this integral at $V_0$ by imposing 
an IR regulator (compactification) on the target space. But we 
also assumed that our RG flows are generated by relevant operators 
which are localized in target space. Hence, the integral over the 
zero mode in the numerator of the matrix \matrixrel\ converges. 
We conclude that the matrix elements ${\cal G}_{ab}$ are of order 
$1/V_0$, and therefore the $C$ function $C(r)$ changes along the 
RG flow only at this sub-leading order. If we remove the cutoff $V_0$ 
then $C(r)$ remains constant---as anticipated in 
the introduction. But we will find that the first correction to 
this constant term in $C(r)$ contains precisely the 
information we are looking for, i.e.\ that it can be used to 
determine the direction of an RG flow. 

\subsec{The $C$ function in sigma model perturbation theory}

In this subsection we present some results from Tseytlin's 
perturbative calculation of the function $C(r)$ (defined in eqs.\ 
\fotta and \zcf) to first order $\apm$. 
We will utilize these results in the next section, to 
demonstrate that spacetime energy decreases along world-sheet 
RG flow. 

Consider the type II sigma model defined by the action\foot{It will be
convenient for us to work with a shifted dilaton $\Phi$ which vanishes 
at infinity. Correspondingly, $\kappa$ is the asymptotic value of the 
gravitational coupling, proportional to $e^{\Phi_0}$. In addition we will set
 all RR-fields to zero. }
\eqn\sigmaaction{
S^{(2)} = 
{1\over2\pi\apm}\int d^2\!z
\left(
\left(G_{ij}(X)+B_{ij}(X)\right)\p X^i\bar\p X^j+
{\apm\over4}\left(\Phi(X)+\Phi_0\right)R^{(2)}
\right)
+{\rm fermions.}
}
If the model is not conformal, then the spacetime fields $G_{ij},
B_{ij},\Phi$ will depend on the renormalization scale $k$. Tseytlin
\tseytlin\ has computed Zamolodchikov's function $C(r)$ for this 
model at first order in the $\apm$ expansion. His results involve 
a `running central charge' $s(k)$, whose scale dependence is converted into
the $r$-dependence of $C(r)$ by a generalized Fourier transform.

To spell out the details we first define $s(k)$. Let us recall that 
the beta functions for the dilaton and metric in sigma model possess
an expansion of the form 
\eqn\betas{\eqalign{
\beta_{ij} &=
\apm\left(
R_{ij} + 2\nabla_i\nabla_j\Phi -
{1\over4}H_{ikl}H_j{}^{kl}
\right) + \CO(\apm^2), \cr
\beta^\Phi &= \apm\left(
-{1\over2}\nabla^2\Phi + (\nabla\Phi)^2 - {1\over24}H^2
\right) + \CO(\apm^2)\ . 
}}
Here we have included in $\beta^\Phi$ the ghost contribution that cancels
the zeroth-order term $d/4$. The spacetime action can be written in terms
of these beta functions:
\eqn\action{\eqalign{
S(k) &= {1\over2\kappa^2\apm}\int
d^d\!x\sqrt{G}e^{-2\Phi}(G^{ij}\beta_{ij}-4\beta^\Phi) \cr
&\sim {1 \over 2\kappa^2} \int d^d\!x\sqrt{G}e^{-2 \Phi}
\left(
R^{(d)}-{1 \over 12}H^2 - 4 (\nabla \Phi)^2 + 4 \nabla^2 \Phi
\right).
}}
From the first line we infer that the spacetime Lagrangian is a linear 
combination of beta functions and hence it vanishes on shell.

Next we define 
\eqn\volm{
V(k) = {1\over2\kappa^2}\int d^d\!x\sqrt{G}e^{-2\Phi},
}
which---by slight abuse of terminology---we will refer to as the 
``volume'' of space.
As in the previous section, we shall give meaning to the divergent 
quantity $V(k)$ by imposing a target space IR cutoff $V_0$. 
Note that, by \betas, 
\eqn\vdot{
kV'(k) = {\apm\over2}S(k).
}
This relation will be useful below in estimating the change in the volume
during the flow.

Tseytlin's `running central charge' $s(k)$ is defined in terms of the 
spacetime effective action and the volume of spacetime by 
\eqn\sk{
s(k) = {3 d \over 2} - {3 \apm S(k) \over 2V(k)}.
}
Note that $s(k)$ is designed to equal the central charge at conformal
points (recall that $S$ vanishes on shell). The Zamolodchikov $C$ function 
can be obtained from $s(k)$ by a generalized Fourier transform,
\eqn\finrest{
C(r) = \int_0^\infty {dk\over k} f(kr)s(k),
}
whose kernel is given by a combination of Bessel functions:
\eqn\fdef{
f(x) = -{1\over2}x^2J_0(x)+{1\over4}x^3J_0'(x)-xJ_0'(x).
}
(it is understood that the integral in eq.\ \finrest\ is computed 
with a damping factor $e^{-\epsilon k}$, where $\ep$ is taken to zero 
at the end of the computation). Equations \finrest, \action, \volm, and 
\fdef\ together constitute Tseytlin's result for $C(r)$ to first order 
in $\apm$ perturbation theory.

The Fourier kernel function \fdef\ is rather complicated, but only 
two simple properties will be of importance to us. Firstly, we have the 
following normalization condition,
\eqn\intf{
\int_0^\infty {dx \over x} f(x)=1,
}
which may be verified either by direct computation or by applying 
\finrest\ to a conformal sigma model in which both $C(r)$ and $s(k)$ 
are constant and equal to the central charge. Secondly, we have 
$f(x) \sim x^4$ as $x \to 0$.

While the detailed expressions we have presented above apply only to
first order in the $\apm$ expansion, the equations of this subsection may 
formally be generalized to exact relations. In particular \sk, \finrest, 
and \fdef\ actually hold as exact relations when the quantities 
$S(k)$ and $V(k)$ are defined by slight generalizations of the first
equation in \action, and equation \vdot. We will not present the
detailed expressions here, but merely note that the exact expression for 
the Lagrangian in $S(k)$ turns out always to be proportional to a linear 
combination of $\beta$ functions, and so vanishes on shell. 
This fact will be important for us in Section 3 below.

\newsec{Target Space Energy Decreases under World-sheet RG Flow}

We now turn to the study of a class of RG flows in non-compact unitary 
sigma models. Specifically, we will combine the explicit computation 
of $C(r)$ in $\apm$ perturbation theory with the Zamolodchikov 
$c$-theorem. We shall thereby be able to obtain all the results 
we listed in the introduction.  

To begin, let us be precise about the assumptions we make. We study RG
flows with the following properties:    
\item{1.} The UV and IR sigma models have the same asymptotic
behavior at spatial infinity. In two spatial dimensions we are able 
to relax this assumption and to allow for different asymptotic conical 
deficit angles in the UV and the IR. 
\item{2.} The RG flow is seeded by a localized tachyon, and proceeds 
by an expanding bubble of the final solution (i.e.\ the IR background) 
embedded in the initial solution (i.e.\ the UV background) with some 
transition region connecting them (see fig.\ 2).%
\foot{Adams, Polchinski, and Silverstein \aps\ first explained that 
tachyon condensation proceeds in this fashion in certain models. 
See Appendix A for an exact solution of the supergravity RG flow 
that illustrates this bubble nucleation and growth in the simple 
context of two dimensional sigma models.} 
\item{3.} All length scales in the asymptotic region of the geometry are
large in string units. This restriction is imposed to permit the use of 
sigma model perturbation theory in the asymptotic region. Note that we do 
not require sigma model perturbation theory to apply uniformly on target 
space: flows involving orbifold fixed points or other singularities are 
perfectly acceptable. Technically, we will apply the $\apm$ expansion 
only in the transition region of fig.\ 2, and only after RG flow has 
proceeded sufficiently far so that the transition region is well 
within the asymptotic regime.
\fig{As RG flow proceeds, a bubble of the
final solution (``Soln 2'') is nucleated and grows. This bubble is
connected to the surrounding sea of the initial solution (``Soln 1'') 
by an interpolating transition region.}{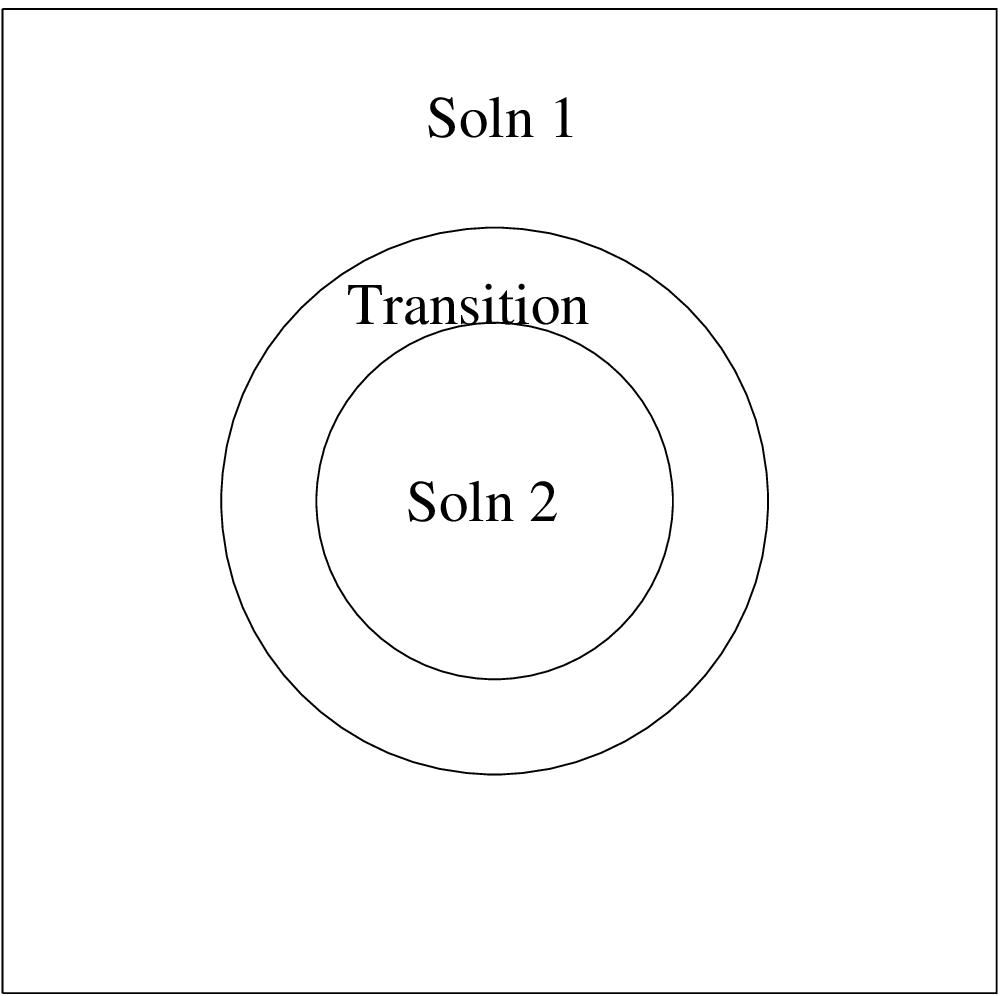}{2.5truein}

\subsec{General strategy}

The main aim of this subsection is to outline the strategy we shall 
follow in order to establish the claims we made in the introduction. 
Some of the necessary but more technical computations will be 
deferred to later subsections. Our argument essentially proceeds
in two steps. In the first step we show that $S(k)$ tends to constant 
for $k \ll k_1$ where $k_1$ is the RG scale at which the expanding 
transition region lies entirely in the asymptotically flat region 
of spacetime. Furthermore, we compute the value of $S(k)$ in this 
infrared region using results from subsection 2.3 and find   
\eqn\deltae{
S(k) \sim -\Delta E_{\rm ADM} 
\equiv E^{\rm IR}_{\rm ADM}- E^{\rm UV}_{\rm ADM},\qquad k \ll k_1, 
 }
where the ADM energies for the IR and UV fixed points are computed 
from the Einstein frame background fields using \energy. Eq.\ \deltae\ 
is established in subsection 3.2--3.4. 

Once we know the behavior \deltae\ of $S(k)$ in the infrared, we can 
pass to the second step. With the help of formulae \finrest, \intf\ 
we will be able to conclude 
\eqn\conclusa{
\Delta E_{\rm ADM} = 
\lim_{V_0\to\infty}{2V_0\over3\alpha'}\Delta C.
}
Here $\Delta C$ represents the difference between the $C$ function 
at very large and very small values of $r$. The formula \conclusa\ 
is the principal result of this paper. According to the $c$-theorem, 
the right hand side of \conclusa\ is negative. Hence we conclude 
that $\Delta E_{\rm ADM}$ is negative, and ADM energy decreases 
under RG flow.

In the remaining part of this subsection we would like to prove eq.\
\conclusa, assuming \deltae\ (which will be proved in later subsections).
We must first admit, however,  that in the paragraphs above 
we have glossed over some important subtleties 
which arise from the IR (in target space) cutoff $V_0$. The latter 
determines the volume of the initial (UV) target space, i.e.\  
$V_0=V(k\!=\!\infty)$. Since this cutoff is merely an artificial 
device, we are not interested in the details of the flow once the 
bubble of fig.\ 2 becomes so large that it begins to sense the IR 
cutoff.\foot{In particular, this allows us to neglect any boundary terms
that might arise due to the IR cutoff, since they will not change during
the part of the flow that interests us.}
Since the RG flow equations are diffusive in character, we
can estimate that this occurs at an RG 
scale $k_0$ determined by
\eqn\kVrel{
-\ln(\apm k_0^2) \sim {V_0^{{2 / d}} \over \apm}
.}
Indeed, the result \deltae\ that we quoted above is strictly valid only 
for $k \gg k_0$, and may more accurately be stated as 
\eqn\deltaea{
S(k) \sim -\Delta E_{\rm ADM},\qquad k_0 \ll k \ll k_1,
}
where, as before, $k_1$ is the finite RG scale at which the 
transition region in fig.\ 2 first lies entirely in the asymptotic 
region of target space geometry.

With these remarks in hand we are prepared to prove eq.\ \conclusa. The idea 
is to insert eq.\ \deltaea\ into eq.\ \finrest\ and thereby relate 
the behavior of $S(k)$ in the infrared to the $C$ function. At first 
sight it seems a bit problematic that we do not know the 
value of $s(k)$ for $k < k_0$ or $k > k_1$. However the $c$-theorem 
and unitarity ensure that it is of unit order. Consequently, the 
contribution to $C(r)$ from the region $0 \leq k \leq k_0$ in the 
integral \finrest\ is of order $r^4 k_0^4$ (recall that $f(x) \sim 
x^4$ at small $x$). For $r \sim \apm^{{1/4}}k_0^{-1/2}$, this 
contribution scales like $\exp({-2{V ^{ {2/ d}} / \apm}})$, and so 
is negligible compared to the dominant effects of order ${1 / V_0}$. 
Similarly, the contribution of $k>k_1$ to the integral \finrest\ is 
of order $(\int_{k_1 r}^{\infty}dx f(x)/x)/V_0 $, and so is
negligible compared to the dominant effect (recall that, from \intf, 
$\lim_{r \to \infty} \int_{k_1 r}^{\infty}dx f(x)/x =0$). 

Based on these estimates it is rather easy to establish \conclusa,
or rather the following more accurate version:
\eqn\conclus{
\Delta E_{\rm ADM} = 
\lim_{V_0\to\infty}{2V_0\over3\alpha'}
\left(C(\apm^{1/4}k_0^{-1/2})-C(0)\right).}
Note that the limit on the right hand side is well defined and 
finite. In fact, according to our discussion in subsection 2.2, 
the $C$ function possesses a universal term that is independent
of the cutoff and a more interesting subleading contribution at 
order $1/V_0$. The first term is removed when we substract the
value of the $C$ function at two different arguments and hence 
the leading non-zero contribution to the difference appears at 
order $1/V_0$. Its coefficient is extracted by multiplying with 
$V_0$ and sending $V_0$ to infinity. This is exactly what we are 
instructed to do on the right hand side of eq.\ \conclus.    

It may have puzzled the reader that we treated the running volume 
$V(k)$ as if it was a constant. This is actually justified because 
(as may be seen using eqs.\ \vdot\ and \kVrel) the fractional change of 
$V(k)$ between the UV and the scale $k_0$ goes to zero in the limit 
of large $V_0$.\foot{This statement actually is only true for $d>2$. For
$d=2$ 
the same fractional change goes to a constant at the scale $k_0$. This
can be derived by considering two truncated cones with the same base 
circumference but different deficit angles. At larger scales such 
as $k_0^{3/4}$, however, the fractional change is zero and 
this is sufficient for the purposes of our argument.}

It remains to prove relation \deltaea.  As we have stressed above, the 
spacetime action $S(k)$ receives no contributions from regions in which 
the field configurations are on shell. Consequently, the action of a 
field configuration of the form depicted in fig.\ 2 receives contributions 
only from the transition region. We now proceed to verify eq.\ \deltaea\ 
by evaluating the action of this transition region for three cases in turn, 
beginning with the most straightforward one, namely asymptotically flat 
backgrounds in $d>2$.

\subsec{Asymptotically flat backgrounds}

Here we assume that both the initial and final target spaces are
asymptotically flat solutions to the Einstein equations. By the 
Newton law, all fields in these solutions deviate at a distance 
$R$ from their flat space values at order $R^{2-d}$. Hence if we 
let the RG flow proceed until the bubble of final solution 
(``Soln 2'' in fig.\  2) is of radius $R$, then field fluctuations 
in the interpolating transition shell are also of order $R^{2-d}$. 
Thus for $d>2$ and at large $R$ all fields in the transition shell 
are very near their flat space values, and so the action of the 
transition band is well approximated by expanding the action in 
powers of the fluctuations of fields. In fact, as $R \to \infty$, 
this expansion need only be performed to linear order; quadratic 
and higher terms in the expansion are negligible. For example, 
the standard quadratic kinetic term for the dilaton, evaluated 
on the transition region, is
\eqn\estimate{
\int d^d\!x \left(\p\Phi\right)^2 = \CO(R^{2-d}),
}
and it may therefore be dropped in the limit $R \to \infty$. 
Higher order terms and terms with more derivatives are further 
suppressed. Indeed the only terms that survive this limit are 
two-derivative terms linear in fluctuations (e.g.\ $\int d^d\!x\,
\p^2\Phi$) which are clearly of order one.

Thus we need only retain the following terms from the action \action:
\eqn\acti{
S_{\rm trans} =
{1\over 2\kappa^2 } \int_{\rm trans} d^d\!x
\sqrt{G}e^{-2 \Phi} \left(R^{(d)}+ 4 \p^2 \Phi \right).
}
Since the ADM energy is defined in terms of the $d+1$ 
dimensional Einstein frame metric,  
\eqn\ef{
g_{00}=-e^{-4 \Phi/(d-1)}, \qquad
g_{ij}=e^{-4 \Phi/(d-1)} G_{ij}, \qquad
g_{0i}=0,
}
we prefer to express eq.\ \acti\ in terms of Einstein frame 
background fields. After dropping quadratic terms the 
result is  
\eqn\actii{
S_{\rm trans} =
{1\over 2\kappa^2 }
\int_{\rm trans} d^d\!x\sqrt{-g^{(d+1)}}R^{(d)}_{\rm E},
}
where $g^{(d+1)}$ is the determinant of the $d+1$ dimensional 
metric \ef, and $R^{(d)}_{\rm E}$ is the 
Ricci scalar formed its spatial part $g_{ij}$.
We can now expand to first order in fluctuations about flat space to obtain
\eqn\actt{
S_{\rm trans} =
{1\over 2\kappa^2 } \int_{\rm trans}d^d\!x\left(\p_i \p_jh_{ij}-\p^2
h_{ii}\right)
= {1\over 2\kappa^2} \int_{\p\rm(trans)} dS^j\left( \p_ih_{ij}- \p_j h_{ii}
\right)
= - \Delta E_{\rm ADM}.
}
The minus sign in the last equality arises because the unit normal 
vector in the integration over the boundary of the transition region 
is directed to the outside of this region. In relating the integrals 
to the definition of the ADM energy we therefore have to invert the 
orientation on the inner shell that bounds the bubble of the final 
solution.

\subsec{Arbitrary asymptotics}

Let us generalize the result of the previous subsection to RG flows in
which the asymptotic boundary conditions shared by the initial and final
solutions may be arbitrary. In such a situation, while an absolute notion
of total energy does not exist, the difference in energies $\Delta E_{\rm
ADM}$ continues to be well defined. For this purpose, we denote the
difference between the two metrics as $h_{ij}$,
\eqn\metdiff{
g_{ij}^{\rm (f)} = g_{ij}^{\rm (i)} + h_{ij},
}
and use the initial metric $g_{ij}^{\rm (i)}$ to raise and lower
indices and calculate connection coefficients. The difference in energies
between the two solutions is then \hh
\eqn\hhen{
\Delta E_{\rm ADM} =
{1 \over 2\kappa^2} \int \sqrt{-g_{00}} (D_i h^{ij} -D^{j} h) dS_{j}.
}
As usual, the integral in \hhen\ is taken over a surface at the boundary
of space at a fixed time, and $S_j$ is an outward-directed unit normal to
the surface. The formula \hhen\ holds for cases in which the metrics are 
independent of time, and the space-time metric components $g_{0i}$ 
vanish. 

We will now show that the action $S(k)$ in the deep IR equals precisely
minus the change \hhen\ in the ADM energy. As in the previous subsection 
this action receives contributions only from the transition shell, and it 
may be evaluated 
by an expansion to first order in fluctuations about the background. 
It is again useful to rewrite the action as a function of the $d+1$
dimensional Einstein frame metric \ef:
\eqn\efeff{
S_{\rm trans} =
{1 \over 2\kappa^2} \int_{\rm trans} d^d\!x \sqrt{-g^{(d+1)}}
\left(R^{(d)}_{\rm E}
+{1\over 12}e^{-8\Phi/(d-1)}H^2 -{4 \over d-1} (\p \Phi)^2
\right).
}
Expanding the action to first order in fluctuations about the background 
we find
\eqn\eftr{
S_{\rm trans} =
{1 \over 2\kappa^2}
\int_{\rm trans} d^d\!x\sqrt{-g^{(d+1)}}g^{ij} \d R_{{\rm E}ij}^{(d)} +
\cdots,
}
where $\cdots$ refers to terms that vanish by the equations of motion on 
the background field. Utilizing the Palatini formula for the variation of 
the Ricci tensor we find
\eqn\actvar{\eqalign{
S_{\rm trans} &= 
{1\over2\kappa^2}\int_{\rm trans}d^d\!x\sqrt{-g^{(d+1)}}
\left( D_i D_j h^{ij} - D^2 h \right) \cr
&= {1 \over 2\kappa^2}
\int_{\p\rm(trans)}\sqrt{-g_{00}}(D_i {h^i}_j -D_j{h^i}_i) dS^j \cr
&= -\Delta E_{\rm ADM}.
}}

\subsec{Analysis for $d=2$}

In 2+1 dimensions, solutions to the Einstein equations that asymptotically
take the form
\eqn\tpo{
ds^2=-dt^2+dr^2+\zeta^2 r^2 d \phi^2
} 
may be assigned an ADM energy proportional to their deficit angle:
\eqn\tde{
E_{\rm ADM} = {2\pi\over\kappa^2}(1-\zeta).
}

Here we shall consider RG flows that interpolate between two solutions 
of the form \tpo\
where the UV deficit angle is $\zeta_1$ and the IR deficit angle is
$\zeta_2$.\foot{In dimensions $d>2$ we 
studied transitions between asymptotically flat backgrounds with 
nontrivial dilaton tails $\Phi \sim {1 / r^{d-2}}$. For $d=2$ such 
a dilaton tail $\Phi \sim \ln(r)$ would destabilize the background  
\tpo. Consequently, we restrict attention to asymptotically
constant dilaton profiles.}
As $H_{ijk}$ is always zero in two dimensions, the string effective 
action \action\ takes an extremely simple form for such flows
\eqn\seat{
S = {1 \over 2\kappa^2} \int d^2\!x \sqrt{-G} R.
}
The integrand on the right hand side is a total derivative. When 
integrated over the transition region it evaluates to
\eqn\stt{\eqalign{
S_{\rm trans} &= {2\pi\over\kappa^2}(\zeta_2-\zeta_1) \cr
&= -\Delta E_{\rm ADM}.
}}
Our ability to control flows that change the asymptotic geometry in two but
not in higher dimensions may be attributed 
to the fact that solutions with distinct deficit angles in two
dimensions differ in energy by a finite amount.

Appendix A provides an illustration of such flows by exhibiting an exact
solution to the RG flow equations at first order in $\alpha'$ that
interpolates between spaces with arbitrary $\zeta_1$ and
$\zeta_2<\zeta_1$.

\newsec{Discussion}

In this note we have studied renormalization group flows that interpolate
between two spacetime solutions with a finite energy difference. These
processes include flows on non-compact sigma models with fixed asymptotics
in $d > 2$,  and flows that interpolate between solutions with different
asymptotic deficit angle in $d=2$. In all these examples we  have argued
that RG flow proceeds in the direction that reduces spacetime energy.

In our proof we have assumed that it is possible to impose an infrared
cutoff on the target space in order to make CFT$_2$ compact. While we do
not have a general proof that such a cutoff always exists, we can
give examples in simple cases. $C/Z_n$ orbifolds can be embedded in compact
$T^2/Z_n$ CFTs for  $n=2,3,4,6$ \VafaRA. For other values of $n$, a
straightforward compactification can be achieved by adding an extra
dimension: consider a level $k$ SU(2) WZW model orbifolded by a $Z_n$
subgroup of SU(2). For large $k$, in the vicinity of the orbifold fixed
line, the target space will approximate $R\times C/Z_n$. The size of the
regulator here is controlled by $k$. It would be interesting---and
useful---to find such regulators for more complicated unstable backgrounds.

It would also be interesting to check explicitly the validity of the
$c$-theorem for the expressions derived in \tseytlin\ for the $C$
function.

Several recent examples of tachyon condensation (including the decay of 
$C^2/Z_n$ orbifolds of type II theory \refs{\aps,\VafaRA,\dv,\chicago}, the 
decay of twisted circle or Melvin compactifications of string theory 
\refs{\dhgm,\DowkerGB,\DowkerSG}, and the decay of non-abelian flux 
backgrounds in the heterotic theory \david), involve more violent 
processes, namely RG flows that change the asymptotics.
The analysis presented in this paper does not 
generalize directly to such flows. The spacetime action $S(k)$ does 
not tend to a constant value in the IR (as in eq.\ \deltae) but instead 
increases without bound.\foot{For example, in flows that interpolate 
between $C^2/Z_n$ and $C^2/Z_m$ one finds that $S(k) \sim -\ln (k 
\sqrt{\apm} )$ by power counting.}   
The $c$-theorem would appear to imply the positivity 
of $S(k)$ at small $k$, but it is not obvious how
to translate this statement into a restriction on RG flows. 

Although we are not able to make this conjecture precise, it is tempting 
to speculate that the violent RG flows discussed in the previous paragraph 
also proceed so as to reduce spacetime energy in some sense.\foot{See
Appendix B for some thoughts in this direction, and \dv\ and \chicago\ for
related proposals.}
Indeed, the most satisfactory completion of the line of argumentation
presented in this paper would be the identification of spacetime energy with 
a natural construction in two dimensional field theory, and a demonstration 
that this quantity decreases along world-sheet RG flows.\foot{A world-sheet 
definition of spacetime energy which is valid on shell
 has been formulated recently in \kraus.} 
Indeed, a solution to the equivalent problem for open strings was proposed 
long ago, and is by now well understood in the context of open string tachyon 
condensation \SenSM. The open string theory partition function on the disk 
(which is related to the boundary entropy in boundary conformal field theory
\AffleckTK) may be identified with spacetime energy and appears to 
decrease along RG flows \aflu. It is possible that a suitably regulated 
version of the closed string partition function on the sphere enters into 
a similar construction for closed strings\foot{We thank E.\ Witten for this
suggestion.} (see \refs{\TseytlinWW,\TseytlinTV} for some work in this
direction). We postpone speculations in this direction to future work.

\bigskip

\centerline{\bf Acknowledgements} We are grateful to A. Adams, T. Banks,
A. Dabholkar, A. Dhar, R. Gopakumar, S. Gukov, P. Kraus, G. Mandal,
E. Martinec, H. Osborn, J. Polchinski, A. Sen,
E. Silverstein, A. Strominger, T. Takayanagi, A. Tseytlin, C. Vafa, 
S. Wadia, and  E. Witten for useful conversations. 
S.M.\ would like to thank the Tata Institute for Fundamental Research, 
Mumbai and A.E.I.\ Golm for hospitality while this work was in progress. 
V.S.\ is grateful for the hospitality of the string group at Harvard. The work 
of M.H.\ and S.M.\ was supported in part by DOE grant 
DE-FG02-91ER40654, and in part by a Harvard Junior Fellowship. 
The work of M.G.\ was supported by NSF grant 9870115.

\appendix{A}{Exact RG Solution for the Decay of a Cone}

The RG flow equation for a sigma model onto a purely gravitational 
target space is, to first order in $\alpha'$,
\eqn\rgeqn{
{dG_{\mu\nu}\over d\lambda} =
-\alpha'R_{\mu\nu} + \nabla_\mu\xi_\nu +\nabla_\nu\xi_\mu,
}
where the vector $\xi$ is arbitrary, representing the freedom to make
continuous changes of target space coordinates along the flow. In \aps\ 
an approximate solution to this equation was presented describing the 
decay of a cone $C/Z_n$ to another one $C/Z_{n'}$ ($n'<n$), for large 
$n$ and $n'$. Here we shall present an exact solution describing the 
decay of a cone from any deficit angle to any smaller deficit angle.

For simplicity let us begin with the case where the cone, with initial
deficit angle $2\pi(1-\zeta)$, decays to the plane. The solution is
\eqn\solone{
ds^2 = \lambda(f^2dr^2+r^2d\phi^2), \qquad
\xi = {1\over2}rfdr,
}
or, by making the $\lambda$-dependent change of coordinates
$r=\rho/\sqrt{\lambda}$,
\eqn\soltwo{
ds^2 = f^2d\rho^2+\rho^2d\phi^2, \qquad
\xi = {1\over2\lambda}\rho f(1-f)d\rho.
}
Here $r,\rho\in[0,\infty)$, $\phi\sim\phi+2\pi\zeta$, and $f$ is a function
that interpolates smoothly between $f=\zeta$ at $r=0$ and $f=1$ at $r=\infty$ 
(see fig.\  3). Specifically,
\eqn\fdef{
f = \left[1+W\left(({1\over\zeta}-1)\exp\left({1\over\zeta}-1-{r^2\over2\alpha'}\right)\right)\right]^{-1},
}
where $W$ is the product log function, the inverse function of $xe^x$.
\fig{The function $f$ defined in \fdef, plotted against $r/\sqrt{\alpha'}$,
for the case $\zeta=1/3$.}{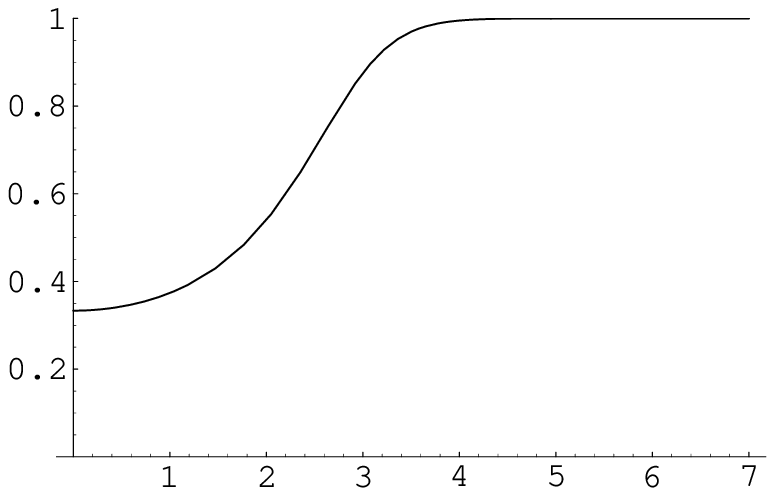}{3truein}

The RG flow parameter $\lambda$ in this solution ranges only from $0$ to
$\infty$. The limits $\lambda\to0$ and $\lambda\to\infty$ are best taken at
fixed $\rho$ (rather than fixed $r$), giving respectively the cone and the
plane. For finite $\lambda$, as shown in fig.\  4, the geometry is conical
at infinity but smooth at the origin, with typical radius of curvature
$\sqrt{\lambda\alpha'}$ (to be precise, the Ricci scalar is
$R=2(f^{-1}-1)/\lambda\alpha'$). Thus the curvature which was initially
concentrated at the origin eventually diffuses over an infinite area. We
can only trust this solution for $\lambda\gg1$, when the curvature is 
small enough for eq.\ \rgeqn\ to be valid. Hence there is no significance to 
the fact that we cannot continue this solution to negative $\lambda$, and 
as usual, we expect the full RG flow to extend from $\lambda=-\infty$ to 
$\lambda=\infty$.

\fig{Cross section of the cone \solone, \soltwo\ in the case
$\zeta=1/3$.}{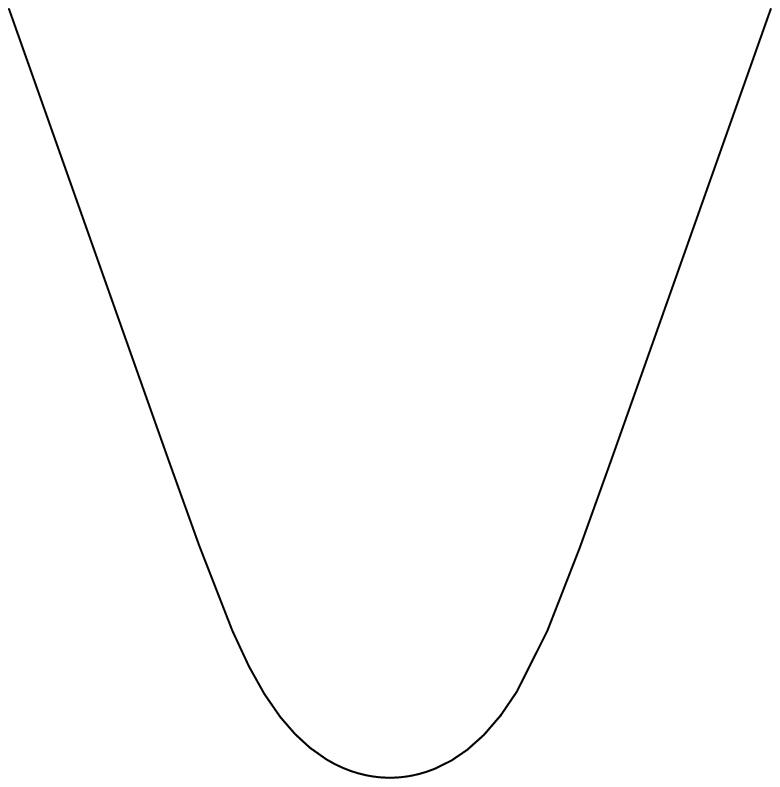}{2.4truein}

The first form \solone\ makes it manifest that the flow proceeds by a 
global Weyl transformation,
i.e.\ the geometry keeps the same shape while expanding in size. This is
analogous to the Gaussian solution of a linear diffusion equation, which over
time broadens but always remains a Gaussian. In this sense a Gaussian is a
``fixed point modulo broadening'' of the diffusion equation, and it is this
property which implies that an arbitrary initial distribution (with a finite
total amount of matter) will after a sufficiently long time diffuse into a
Gaussian distribution. Analogously, since the above solution is a ``fixed
point modulo global Weyl transformations'' of eq.\ \rgeqn, we can expect 
that if at $\lambda\approx1$ the geometry consists of a cone with a 
smoothed-out 
vertex of size $\sim\sqrt{\alpha'}$, then for large $\lambda$ the flow 
will be better and better approximated by the solutions \solone\ and 
\soltwo, with 
errors of order $1/\lambda$. This is important because it implies that, 
in studying the RG flow of a $C/Z_n$ orbifold, it is not necessary to 
know the precise configuration at the end of the string-scale regime in 
order to know how the flow will proceed in the subsequent supergravity 
regime.

The solution can easily be generalized to describe the decay of a cone with
$\zeta=\zeta_1$ to one with $\zeta=\zeta_2>\zeta_1$. Simply let the 
periodicity of $\phi$ be $2\pi\zeta_1$, and set $\zeta=\zeta_1/\zeta_2$ in
the expression \fdef\ for $f$.

\appendix{B}{Speculations on Asymptotics-changing RG Flows}

In this paper we have argued that spacetime energy decreases along
RG flows for which the asymptotic geometry of spacetime remains
constant along flows. Though it is tempting to speculate that this result 
applies to more general RG flows, it is difficult to formulate 
a precise conjecture along these lines, as it is unclear how the energies 
of solutions with different asymptotics may be compared.

In this connection, however, we find it intriguing to note that Hawking 
and Horowitz have derived a rather beautiful universal and geometrical 
formula \hh 
\eqn\enhh{
E_{\rm HH}=  - {1 \over 8\pi} \int \sqrt{-g_{00}} K
}
for the energy of spacetimes of the form
\eqn\stm{
ds^2= g_{00}dt^2+g_{ij}dx^i dx^j
}
where the integral is taken over the boundary of space at a fixed time, 
and $K$ is the extrinsic curvature of this boundary viewed as a sub-manifold 
of a constant time slice of the metric \stm. Unfortunately, the relationship 
\enhh\ is rather formal since the right hand side diverges for most 
spacetimes. 
To arrive at a finite quantity one again has to subtract the contribution of 
a reference background in \enhh. 

Nonetheless eq.\ \enhh\ is useful and appears to fit well with world-sheet RG 
flows. It was used in \hh\ to derive the formula \hhen\ for the difference 
in energy between two nearby spacetimes. It is conceivable that eq.\ \enhh\ 
may also be employed, together with a regulation prescription motivated by a 
consideration of RG flows, to determine the sign of the (generically 
divergent) energy differences between spacetimes of different asymptotics.

To illustrate the idea, consider an RG flow from a ${C^2 / Z_n}$
orbifold to a ${ C^2 / Z_m}$ orbifold of type II theory.  The
extrinsic curvature of a sphere of radius $R$ is $1/R$. Consequently, 
the energy \enhh\ of the ${C^2 / Z_n}$, computed on a sphere of radius 
$R_1$, is  $-R_1^2/8 \pi n$. Similarly the energy of the ${C^2 / Z_m}$ 
orbifold computed on a sphere of radius $R_2$ is ${-R_2^2 /8 \pi  m}$. 
As a toy model of an RG flow motivated regulation procedure, we prescribe 
that the energies of $C^2/Z_n$ and $C^2/Z^m$ should be compared on 
surfaces of equal surface area, i.e.\ when $R_1^3/n=R_2^3/m$. With this 
prescription,
\eqn\enhh{
\Delta E = 
{R_1^2 \over 8 \pi n}\left(1-\left({n \over m}\right)^{1/3}\right).
}
If we conjecture that energy decreases along RG flows then we derive a 
constraint on these flows: The order of the orbifold must decrease in 
the process of tachyon condensation. This prediction happens to be
fulfilled in all examples of such RG flows that we are aware of.
Nonetheless the analysis of this paragraph should not be taken too 
seriously. It is presented merely as a cartoon of an detailed 
analysis which remains to be worked out and may succeed in using 
an energy function like \enhh\ to constrain RG flows.

\listrefs

\end